\documentclass[conference]{IEEEtran}
\ifCLASSINFOpdf
\else
\fi

\usepackage{graphicx,color}				
\usepackage{amssymb,amsmath,amsfonts,amsthm}  % Math packages

\usepackage{algorithm,caption}
\usepackage[noend]{algpseudocode}

\newcounter{MYtempeqncnt}

\def \tr {\text{tr}}

\begin{document}
%
% paper title
% Titles are generally capitalized except for words such as a, an, and, as,
% at, but, by, for, in, nor, of, on, or, the, to and up, which are usually
% not capitalized unless they are the first or last word of the title.
% Linebreaks \\ can be used within to get better formatting as desired.
% Do not put math or special symbols in the title.
%\title{FEC Code Anchored MIMO Receiver\\ Using Semi-definite Relaxation}
\title{Non-iterative Joint Detection-Decoding Receiver for LDPC-Coded MIMO Systems Based on SDR}

% author names and affiliations
% use a multiple column layout for up to three different
% affiliations
% =========== for conference ==============
\author{\IEEEauthorblockN{Kun Wang}
\IEEEauthorblockA{Qualcomm Technologies, Inc.\\
3165 Kifer Road \\
Santa Clara, CA 95051, USA \\
Email: kunwang@ieee.org}
\and
\IEEEauthorblockN{Zhi Ding}
\IEEEauthorblockA{Dept. of Electrical and Computer Engineering\\
University of California, Davis\\
Davis, CA 95616, USA\\
Email: zding@ucdavis.edu}
}

\maketitle

% As a general rule, do not put math, special symbols or citations
% in the abstract
\begin{abstract}
Semi-definite relaxation (SDR) detector has been demonstrated to be successful in 
approaching maximum likelihood (ML) performance while the time complexity is only polynomial. 
We propose a new receiver jointly utilizing the forward error correction (FEC) code information
 in the SDR detection process.
Strengthened by the code constraints, the joint SDR detector provides soft information
of much improved reliability to downstream decoder and therefore outperforms 
existing receivers with substantial gain.
\end{abstract}

% no keywords

% For peer review papers, you can put extra information on the cover
% page as needed:
% \ifCLASSOPTIONpeerreview
% \begin{center} \bfseries EDICS Category: 3-BBND \end{center}
% \fi
%
% For peerreview papers, this IEEEtran command inserts a page break and
% creates the second title. It will be ignored for other modes.
\IEEEpeerreviewmaketitle

\section{Introduction}  \label{sec:intro}

% MIMO advantage
Multiple-input multiple-output (MIMO) transceiver
technology represents a breakthrough in the advances of 
wireless communication systems. 
Modern wireless systems widely adopt multiple antennas, for example, the 3GPP LTE and WLAN systems \cite{molisch2012wireless},
and further massive MIMO has been proposed for next-generation wireless systems \cite{larsson2014massive}.
MIMO systems can provide manifold throughput increase, or can offer reliable transmissions by spatial diversity \cite{goldsmith2005wireless}. 
In order to fully exploit the advantages promised by MIMO, the 
receiver must be able to effectively recover the transmitted information.
Thus, detection and decoding remain to be one of the fundamental areas in state-of-the-art MIMO research.

% ML and sub-optimal receivers
It is well known that maximum likelihood (ML) detection is optimal in terms 
of minimum error probabilities for equally likely data sequence transmissions. 
%when \textit{a priori} information is unavailable. 
However, the ML detection is NP-hard \cite{verdu1989computational} and its time complexity is exponential for MIMO detection, regardless of whether exhaustive search or other search algorithms (e.g., sphere decoding) 
are used \cite{jalden2005complexity} in data symbol detection. 
Aiming to reduce the high computational complexity for MIMO receivers,
a number of research efforts have focused on designing
near-optimal and high performance receivers. In the literature, 
the simplist linear receivers, such as matched filtering (MF), zero-forcing (ZF) and minimum mean squared error (MMSE), have been widely investigated. 
Other more reliable and more sophisticated receivers, such as
successive interference cancellation (SIC) or
parallel interference cancellation (PIC) receivers have also been
studied.  However, these receivers suffer substantial performance loss. 

% SDR advantage
In recent years, various semi-definite relaxation techniques have
emerged as a sub-optimum detection method that 
can achieve near-ML detection 
performance \cite{luo2010semidefinite}.
Specifically, ML detection of MIMO transmission 
can be formulated as least squares integer programming problem 
which can then be converted into an 
equivalent quadratic constrained quadratic program (QCQP). 
The QCQP can be transformed  by 
relaxing the rank-1 constraint into a semi-definite program.  
With the name \textit{semi-definite relaxation} (SDR), 
its substantial performance improvement over algorithms such as
MMSE and SIC has stimulated broad research interests as seen in
the works of
\cite{tan2001application,ma2002quasi,ma2004semidefinite,ma2006blind}. 
Several earlier works \cite{tan2001application,ma2002quasi} developed 
SDR detection in proposing multiuser detection for CDMA transmissions. 
Among them, the authors of \cite{ma2004semidefinite} proposed an SDR-based 
multiuser detector for $M$-ary PSK signaling. Another work in
\cite{ma2006blind} presented an efficient SDR 
implementation of blind ML detection 
of signals that utilize orthogonal space-time block codes. 
Furthermore, multiple SDR detectors of 16-QAM signaling 
were compared and shown to be equivalent in \cite{ma2009equivalence}.

% Our contribution
Although most of the aforementioned studies focused on SDR 
detections of uncoded transmissions, forward error correction (FEC)
codes in binary field have long been integrated into data communications
to effectively combat noises and co-channel interferences. 
Because FEC decoding takes place in the finite binary field whereas  
modulated symbol 
detection is formulated in the Euclidean space of complex field, 
the joint detection and decoding typically relies on the concept of
iterative turbo processing. 
In this work, however, we present a non-iterative receiver based on SDR 
for joint detection and decoding.
In our design, FEC codes not only are used for decoding, but also are
integrated as constraints 
within the detection optimization formulation to 
develop a novel joint SDR detector
\cite{wang2014joint,wang2016robust,wang2015diversity,wang2017galois}. 
Instead of using the more traditional randomization or rank-one 
approximation for symbol detection, our data detection takes
advantage of the last column of the optimal SDR matrix solution. 
When compared with the original SDR detector in \cite{luo2010semidefinite},
our integrated SDR receiver demonstrates substantial performance gain.

%This manuscript is organized as follows. 
%First, Section \ref{sec:sys_model} describes the baseband MIMO system model 
%and a corresponding real field detection problem. 
%We also present the SDR formulation to approximate the maximum likelihood
%detection of MIMO signals.
%In Section \ref{sec:code_anchor}, we integrate
%the FEC code constraints into the SDR to form a joint ML-SDR receiver.
%We demonstrate the superior performance of joint SDR receiver in Section \ref{sec:sim}.
%Finally, Section \ref{sec:con} concludes the work.

\section{System Model and SDR Detection}  \label{sec:sys_model}

\subsection{Maximum-likelihood MIMO Signal Detection}
Consider an $N_t$-input $N_r$-output spatial multiplexing MIMO system with memoryless channel.
The baseband equivalent model of this system at time $k$ can be expressed as
\begin{equation} \label{eq:mimo_complex}
\mathbf{y}_k^c = \mathbf{H}_k^c \mathbf{s}_k^c + \mathbf{n}_k^c, \quad k = 1, \ldots, K,
\end{equation}
where $\mathbf{y}_k^c \in \mathbb{C}^{N_r \times 1}$ is the received signal,
$\mathbf{H}_k^c \in \mathbb{C}^{N_r \times N_t}$ denotes the MIMO channel matrix,
$\mathbf{s}_k^c \in \mathbb{C}^{N_t \times 1}$ is the transmitted signal, and 
$\mathbf{n}_k^c  \in \mathbb{C}^{N_r \times 1}$ is an additive Gaussian noise vector, 
each element of which is independent and follows $\mathcal{CN}(0, 2 \sigma_n^2)$. 
In fact, besides modeling the point-to-point MIMO system, Eq.~(\ref{eq:mimo_complex}) can be also
used to model frequency-selective systems 
\cite{lathi2009modern}, multi-user systems \cite{wang2016fec}, among others. 
The only difference lies in the structure of channel matrix  $\mathbf{H}_k^c$.

To simplify problem formulation, the complex-valued
signal model can be transformed into
the real field by letting
\begin{equation*}
\mathbf{y}_k = 
\begin{bmatrix}
\text{Re}\{ \mathbf{y}_k^c  \} \\
\text{Im}\{ \mathbf{y}_k^c  \}
\end{bmatrix},
\mathbf{s}_k = 
\begin{bmatrix}
\text{Re}\{ \mathbf{s}_k^c  \} \\
\text{Im}\{ \mathbf{s}_k^c  \}
\end{bmatrix},
\mathbf{n}_k = 
\begin{bmatrix}
\text{Re}\{ \mathbf{n}_k^c  \} \\
\text{Im}\{ \mathbf{n}_k^c  \}
\end{bmatrix},
\end{equation*}
and
\begin{equation*}
\mathbf{H}_k = 
\begin{bmatrix}
\text{Re}\{ \mathbf{H}_k^c  \}  & - \text{Im}\{ \mathbf{H}_k^c  \}\\
\text{Im}\{ \mathbf{H}_k^c  \} & \text{Re}\{ \mathbf{H}_k^c  \}
\end{bmatrix}.
\end{equation*}
Consequently, the transmission equation is given by
\begin{equation} \label{eq:mimo_real}
\mathbf{y}_k = \mathbf{H}_k \mathbf{s}_k + \mathbf{n}_k, \quad k = 1, \ldots, K.
\end{equation}
In this study, we choose capacity-approaching LDPC code for the purpose of forward error correction. 
Further, we assume the transmitted symbols are generated based on QPSK constellation,
i.e., $s_{k,i}^c \in \{ \pm1 \pm j \}$ for $k = 1, \ldots, K$ and $i = 1, \ldots, N_t$.
The codeword (on symbol level) is placed first along the spatial dimension
and then along the temporal dimension.

Before presenting the code anchored detector, we begin with a brief review of existing 
SDR detector in uncoded MIMO systems for the convenience of subsequent integration.
By the above assumption of Gaussian noise, it can be easily shown that the optimal ML detection
is equivalent to the following discrete least squares problem
\begin{equation} \label{eq:ml_detection}
\underset{\mathbf{x}_k \in \{ \pm 1 \}^{2N_t}}{\text{min.}} \;
\sum_{k=1}^K \Vert \mathbf{y}_k - \mathbf{H}_k \mathbf{x}_k \Vert^2.
\end{equation}
However, this problem is NP-hard. Brute-force solution would take exponential time (exponential in $N_t$).
Sphere decoding was proposed for efficient computation of ML problem. 
Nonetheless, it is still exponentially complex, even on average sense \cite{jalden2005complexity}.

\subsection{SDR MIMO Detector}  \label{sec:uncoded_sdr}
SDR can generate an \textit{approximate} solution to the ML problem
in polynomial time. More specifically, the time complexity is $\mathcal{O}(N_t^{4.5})$ when a generic interior-point algorithm is used,
and it can be as low as $\mathcal{O}(N_t^{3.5})$ with a customized algorithm \cite{luo2010semidefinite}.
The trick of using SDR is to firstly turn the ML detection into a homogeneous QCQP 
by introducing auxiliary variables $\{t_k, k = 1, \ldots, K\}$ \cite{luo2010semidefinite}.
The ML problem can then be equivalently written as the following QCQP
\begin{equation} \label{eq:qcqp}
\begin{aligned}
& \underset{\{\mathbf{x}_k, t_k\}}{\text{min.}}
& &  \sum_{k=1}^K 
\begin{bmatrix}
\mathbf{x}_k^T & t_k
\end{bmatrix}
\begin{bmatrix}
\mathbf{H}_k^T \mathbf{H}_k & \mathbf{H}_k^T \mathbf{y}_k  \\
-\mathbf{y}_k^T \mathbf{H}_k & || \mathbf{y}_k ||^2 
\end{bmatrix} 
\begin{bmatrix}
\mathbf{x}_k \\
t_k
\end{bmatrix} \\
& \text{s.t.}
& & t_k^2 = 1, \; x_{k,i}^2 = 1, \; k = 1, \ldots, K, i = 1, \ldots, 2N_t.
\end{aligned}
\end{equation}

This QCQP is non-convex because of its quadratic equality constraints. 
To solve it approximately via SDR, define the rank-1 semi-definite matrix
\begin{equation} \label{eq:rank1_matrix}
\mathbf{X}_k = 
\begin{bmatrix}
\mathbf{x}_k \\
t_k
\end{bmatrix}
\begin{bmatrix}
\mathbf{x}_k^T & t_k
\end{bmatrix}
=
\begin{bmatrix}
\mathbf{x}_k \mathbf{x}_k^T & t_k \mathbf{x}_k \\
t_k \mathbf{x}_k^T & t_k^2
\end{bmatrix},
\end{equation}
and for notational convenience, denote the cost matrix by
\begin{equation} \label{eq:cost_matrix}
\mathbf{C}_k = 
\begin{bmatrix}
\mathbf{H}_k^T \mathbf{H}_k & \mathbf{H}_k^T \mathbf{y}_k  \\
-\mathbf{y}_k^T \mathbf{H}_k & || \mathbf{y}_k ||^2
\end{bmatrix}.
\end{equation}
Using the property of trace $\mathbf{v}^T\mathbf{Q}\mathbf{v} = \tr(\mathbf{v}^T\mathbf{Q}\mathbf{v}) = \tr(\mathbf{Q}\mathbf{v}\mathbf{v}^T)$,
the QCQP in Eq.~(\ref{eq:qcqp}) can be relaxed to SDR by removing the rank-1 constraint on $\mathbf{X}_k$.
Therefore, the SDR formulation is
\begin{equation} \label{eq:disjoint_sdr}
\begin{aligned}
& \underset{\{\mathbf{X}_k\}}{\text{min.}}
& &  \sum_{k=1}^K \tr(\mathbf{C}_k \mathbf{X}_k) \\
& \text{s.t.}
& & \tr(\mathbf{A}_i \mathbf{X}_k) = 1, \; k = 1, \ldots, K, i = 1, \ldots, 2N_t + 1, \\
& 
& & \mathbf{X}_k \succeq 0, \; k = 1, \ldots, K,
\end{aligned}
\end{equation}
where $\mathbf{A}_i$ is a zero matrix except that the $i$-th position on the diagonal is 1,
so $\mathbf{A}_i$ is used for extracting the $i$-th element on the diagonal of $\mathbf{X}_k$.
It is noted that $\mathbf{A}_i  \equiv \mathbf{A}_{i,k}, \forall k$; thus, the index $k$ is omitted
for $\mathbf{A}_{i,k}$ in Eq.~(\ref{eq:disjoint_sdr}). 
Finally, we would like to point out that the SDR problems formulated in most papers 
are targeted at a single time snapshot, since their system of interest is uncoded.
Here, for subsequent integration of code information,
we consider a total of $K$ snapshots that can accommodate an FEC codeword.

% ############ Double-column equations ###############
% Reference: Page 12 
% http://ctan.math.washington.edu/tex-archive/macros/latex/contrib/IEEEtran/IEEEtran_HOWTO.pdf
\begin{figure*}[!t]
% ensure that we have normalsize text
\normalsize
% Store the current equation number.
\setcounter{MYtempeqncnt}{\value{equation}}
% Set the equation number to one less than the one
% desired for the first equation here.
% The value here will have to changed if equations
% are added or removed prior to the place these
% equations are referenced in the main text.
\setcounter{equation}{11}
\begin{equation} \label{eq:joint_ml_sdr}
\begin{aligned}
& \underset{\{\mathbf{X}_k, f_n\}}{\text{min.}}
& &  \sum_{k=1}^K \tr(\mathbf{C}_k \mathbf{X}_k) \\
& \text{s.t.}
& & \tr(\mathbf{A}_i \mathbf{X}_k) = 1, \, \mathbf{X}_k \succeq 0, \quad k = 1, \ldots, K, i = 1, \ldots, 2N_t+1, \\
&
& & \tr(\mathbf{B}_i \mathbf{X}_k) = 1 - 2 f_{2N_t(k-1)+2i-1}, \quad k = 1, \ldots, K, i = 1, \ldots, N_t, \\
&
& & \tr(\mathbf{B}_{i+N_t} \mathbf{X}_k) = 1 - 2 f_{2N_t(k-1)+2i}, \quad k = 1, \ldots, K, i = 1, \ldots, N_t, \\
&
& & \sum_{ n \in \mathcal{F} } f_n - \sum_{ n \in \mathcal{N}_m \backslash \mathcal{F}} f_n \leq |\mathcal{F}| - 1, \quad \forall m \in \mathcal{M}, \forall \mathcal{F} \in \mathcal{S};\\ 
& &&
0 \leq f_n \leq 1, \quad \forall n \in \mathcal{N}.
\end{aligned}
\end{equation}
% Restore the current equation number.
\setcounter{equation}{\value{MYtempeqncnt}}
% IEEE uses as a separator
\hrulefill
% The spacer can be tweaked to stop underfull vboxes.
\vspace*{4pt}
\end{figure*}
% #########################################

\section{FEC Codes in Joint SDR Receiver Formulation}  \label{sec:code_anchor}
If MIMO detector can provide more accurate 
information to downstream decoder, 
an improved decoding performance can be expected.
With this goal in mind, we propose to use FEC code information when performing detection.

\subsection{FEC Code Anchoring}
Consider an $(N_c,K_c)$ LDPC code. Let $\mathcal{M}$ and $\mathcal{N}$ be the 
index set of check nodes and variable nodes of the parity check matrix, respectively, i.e.,
$\mathcal{M} = \{1, \ldots, N_c-K_c \}$ and $\mathcal{N} = \{1, \ldots, N_c \}$.
Denote the neighbor set of the $m$-th check node as $\mathcal{N}_m$
and let $\mathcal{S} \triangleq \{ \mathcal{F} \, | \, \mathcal{F} \subseteq \mathcal{N}_m \, \text{with} \, |\mathcal{F}| \, \text{odd} \}$.
Then one characterization of fundamental polytope is captured by 
the following forbidden set (FS) constraints \cite{feldman2005using}
\begin{equation} \label{eq:parity_ineq}
\sum_{ n \in \mathcal{F} } f_n - \sum_{ n \in \mathcal{N}_m \backslash \mathcal{F}} f_n \leq |\mathcal{F}| - 1, \; \forall m \in \mathcal{M},
\forall \mathcal{F} \in \mathcal{S}
\end{equation}
plus the box constraints for bit variables
\begin{equation} \label{eq:box_ineq}
 0  \leq f_n \leq 1, \quad \forall n \in \mathcal{N}.
\end{equation}

Recall that the bits $\{f_n\}$ are mapped by modulators into
transmitted data symbols in $\mathbf{x}_k$. 
It is important to note that the parity check 
inequalities (\ref{eq:parity_ineq}) can help to tighten our 
detection solution of $\mathbf{x}_k$
by explicitly forbidding the bad configurations 
of $\mathbf{x}_k$ that are inconsistent with FEC codewords. 
Thus, a joint detection and decoding algorithm can take advantage
of these linear constraints by integrating them within the
SDR problem formualtion. 

Notice that coded bits $\{f_n\}$ are in fact binary. Hence, 
the box constraint of (\ref{eq:box_ineq}) is a relaxation of 
the binary constraints. 
In fact, if variables $f_n$'s are forced to be only 0's and 1's
(binary),  then the constraints (\ref{eq:parity_ineq}) will be equivalent 
to the original binary parity-check constraints.
To see this, if parity check node $m$ fails to hold, 
there must be a subset of variable nodes
$\mathcal{F} \subseteq \mathcal{N}_m$ of odd cardinality 
such that all nodes in $\mathcal{F}$
have the value 1 and all those in 
$\mathcal{N}_m \backslash \mathcal{F}$ have value 0.
Clearly, the corresponding parity inequality in (\ref{eq:parity_ineq}) 
would forbid such outcome.

\subsection{Symbol-to-Bit Mapping}
To anchor the FS constraints into the SDR formulation in Eq.~(\ref{eq:disjoint_sdr}),
we need to connect the bit variables $f_n$'s with the data vectors
$\mathbf{x}_k$'s or the matrix variables $\mathbf{X}_k$'s.

As stated in \cite{luo2010semidefinite}, if $(\mathbf{x}_k^*, t_k^*)$ is an optimal solution to (\ref{eq:disjoint_sdr}),
then the final solution should be $t_k^* \mathbf{x}_k^*$, where $t_k^*$ controls the sign of the symbol. 
In fact, Eq.~(\ref{eq:rank1_matrix}) shows that
the first $2 N_t$ elements of last column or last row are exactly $t_k \mathbf{x}_k$.
We also note that the first $N_t$ elements correspond to the real parts of the transmitted symbols
and the next $N_t$ elements correspond to the imaginary parts. 
Hence, for QPSK modulation,
the mapping constraints
for time instant $k = 1, \ldots, K$ are simply as follows
\begin{equation} \label{eq:qpsk_gray}
\begin{split}
& \tr(\mathbf{B}_i \mathbf{X}_k) = 1 - 2 f_{2N_t(k-1)+2i-1}, \; i = 1, \ldots, N_t, \\
& \tr(\mathbf{B}_{i+N_t} \mathbf{X}_k) = 1 - 2 f_{2N_t(k-1)+2i}, \; i = 1, \ldots, N_t,
\end{split}
\end{equation}
where $\mathbf{B}_i$ is a selection matrix
designed to extract the $i$-th element on the last column/row of $\mathbf{X}_k$ (except last element):
\begin{equation}
\mathbf{B}_i = 
\begin{bmatrix}
0 & \ldots & \ldots & \ldots & 1/2 \\
\vdots   & \ddots &  & & \vdots  \\
\vdots   & & 0 &  & 1/2 \\
\vdots  & & & \ddots & \vdots \\
1/2 & \ldots & 1/2 & \ldots & 0
\end{bmatrix}, \; 1 \leq i \leq 2 N_t.
\end{equation}
The non-zero entry of $\mathbf{B}_i$ is the $i$-th element on the last column. 
For the same reason as that of $\mathbf{A}_i$,
the index $k$ is omitted in $\mathbf{B}_i$.
Moreover, note the subtle difference 
that $\mathbf{A}_i$ is defined for $1 \leq i \leq 2N_t + 1$ 
while $\mathbf{B}_i$ is defined for $1 \leq i \leq 2N_t$.

\subsection{Joint ML-SDR Receiver}
Having defined the necessary notations and constraints, 
a joint ML-SDR detector can be formulated as the optimization
problem in Eq.~(\ref{eq:joint_ml_sdr}) for QPSK modulation.
For higher order QAM beyond QPSK, the necessary changes 
for our joint SDR receiver include the relaxed box constraints for diagonal elements \cite{ma2009equivalence}
and the symbol-to-bit mapping constraints.
We refer interested readers to the works \cite{wang2016fec,wang2015joint,wang2016diversity,wang2017galois} 
for the details of mapping higher order QAM constraints.
%\begin{equation} \label{eq:joint_ml_sdr}
%\begin{aligned}
%& \underset{\{\mathbf{X}_k, f_n\}}{\text{min.}}
%& &  \sum_{k=1}^K \tr(\mathbf{C}_k \mathbf{X}_k) \\
%& \text{s.t.}
%& & \tr(\mathbf{A}_i \mathbf{X}_k) = 1, \, \mathbf{X}_k \succeq 0, \quad k = 1, \ldots, K, i = 1, \ldots, 2N_t+1, \\
%&
%& & \tr(\mathbf{B}_i \mathbf{X}_k) = 1 - 2 f_{2N_t(k-1)+2i-1}, \quad k = 1, \ldots, K, i = 1, \ldots, N_t, \\
%&
%& & \tr(\mathbf{B}_{i+N_t} \mathbf{X}_k) = 1 - 2 f_{2N_t(k-1)+2i}, \quad k = 1, \ldots, K, i = 1, \ldots, N_t, \\
%&
%& & \sum_{ n \in \mathcal{F} } f_n - \sum_{ n \in \mathcal{N}_m \backslash \mathcal{F}} f_n \leq |\mathcal{F}| - 1, \quad \forall m \in \mathcal{M}, \forall \mathcal{F} \in \mathcal{S};\\ 
%& &&
%0 \leq f_n \leq 1, \quad \forall n \in \mathcal{N}.
%\end{aligned}
%\end{equation}

Recall that the matrix $\mathbf{X}_k\succeq 0 $ is a relaxation of
the rank one matrix 
\[
\mathbf{X}_k = 
\begin{bmatrix}
	\mathbf{x}_k \\
	t_k
\end{bmatrix}
\begin{bmatrix}
	\mathbf{x}_k^T & t_k
\end{bmatrix}\]
After obtaining the optimal solution $\{ \mathbf{X}_k \}$ of the SDR, one must
determine the final detected symbol values in $\mathbf{x}_k$. 
Traditionally, one
``standard'' approach to retrieve the final solution
% $(\mathbf{x}_k^*, t_k^*)$ 
is via Gaussian randomization that views $\mathbf{X}_k$ as the covariance matrix
of $\mathbf{x}_k$, and another method is to apply
rank-one approximation of $\mathbf{X}_k$ \cite{luo2010semidefinite}.

However, a more convenient way is to directly use the first $2N_t$ elements in the last column of $\mathbf{X}_k$.
If hard-input hard-output decoding algorithm (such as bit flipping) is used, 
we can first quantize $t_k^* \mathbf{x}_k^*$ into binary values before feeding 
them to the FEC decoder for error correction. 
On the other hand, for soft-input soft-output decoder 
such as sum-product algorithm (SPA), log-likelihood ratio (LLR) can be generated 
from the unquantized $t_k^* \mathbf{x}_k^*$.
%Here, we caution that the unquantized results from Gaussian randomization are not suitable for soft decoders such as the SPA, because the magnitudes of the
%LLRs generated from randomization do not accurately reflect the data bits' 
%actual reliability. 

%\textcolor{red}{Proposition: ML-certificate.}
%\textcolor{blue}{Bit vector $\mathbf{f}$ is integral $\Leftrightarrow \mathbf{X}_k$ is rank-1 $\Leftrightarrow$ ML codeword.}

\section{Simulation Results} \label{sec:sim}
In the simulation tests, a MIMO system with $N_t = 4$ and $N_r = 4$ is assumed.
The MIMO channel coefficients are assumed to be ergodic Rayleigh fading.
QPSK modulation is used and a regular (256,128) LDPC code with column weight 3 is employed.

In this section, we will demonstrate the power of code anchoring. 
We term the formulation in Eq.~(\ref{eq:disjoint_sdr}) as \textit{disjoint ML-SDR}, 
while that in Eq.~(\ref{eq:joint_ml_sdr}) as \textit{joint ML-SDR}.
With the optimal SDR solution $\{\mathbf{X}_k^*\}$, there are several approaches to retrieve 
the final solution $\mathbf{\hat{s}}_k$.
\begin{enumerate}
\item[-] \textit{Rank-1 approximation}: Perform eigen-decomposition on $\mathbf{X}_k^*$ to obtain the largest eigenvalue $e_k$ and its corresponding eigenvector $\mathbf{v}_k$. The final solution $\mathbf{\hat{s}}_k = \sqrt{e_k} \mathbf{v}_k[1:2N_t] \times \mathbf{v}_k[2N_t+1]$.
\item[-] \textit{Direct approach}: The final solution is retrieved from the last column of $\mathbf{X}_k$, 
i.e., $\mathbf{\hat{s}}_k = \mathbf{X}_k[1:2N_t, 2N_t+1]$.
\item[-] \textit{Randomization}: Generate $\mathbf{v}_k \sim \mathcal{CN}(\mathbf{0}, \mathbf{X}_k)$ for a certain number of trials, and pick the one that results in smallest cost value. Note that when evaluating the cost value, the elements of $\mathbf{v}_k$ are quantized to $\{-1,+1\}$.  
\end{enumerate}

\begin{figure}[!htb]
\centering
\centerline{\includegraphics[width=8.75cm]{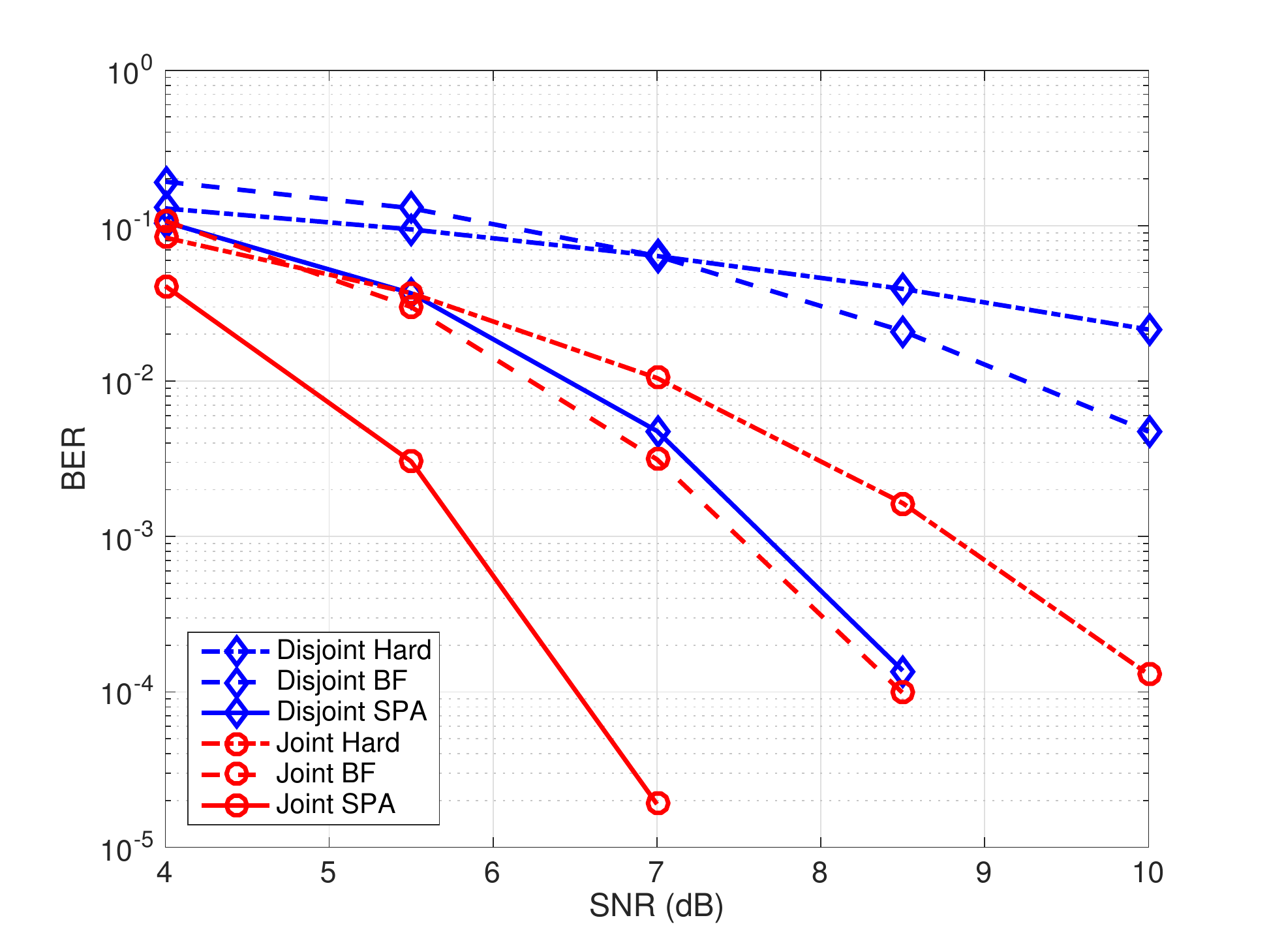}}
\caption{\small{BER comparisons of disjoint and joint SDR receivers: Rank 1 approximation.}}
\vspace*{-3mm}
\label{fig:rank1}
\end{figure}

\begin{figure}[!htb]
\centering
\centerline{\includegraphics[width=8.75cm]{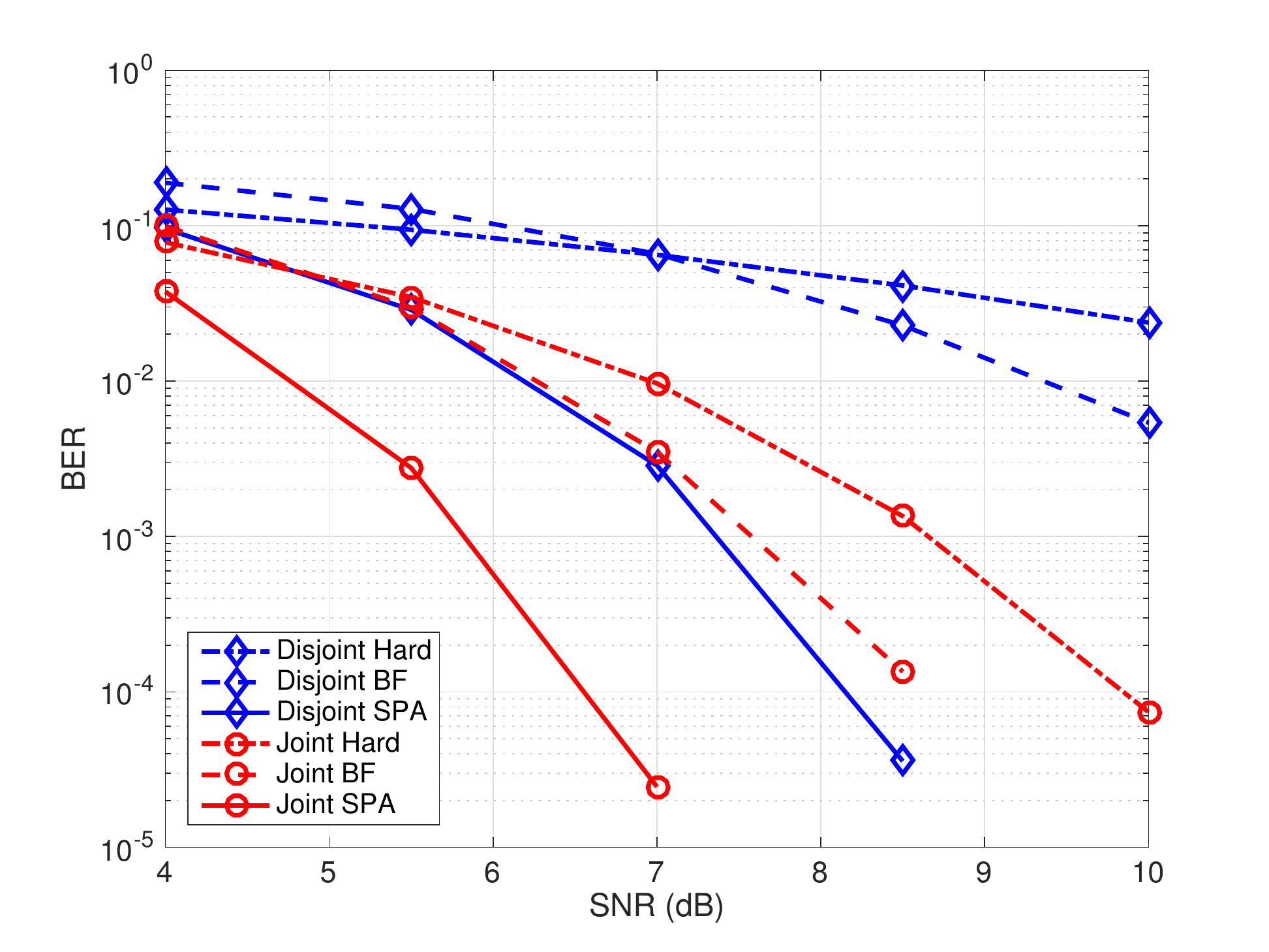}}
\caption{\small{BER comparisons of disjoint and joint SDR receivers: Direct
		approach using the final column of $\mathbf{X}_k$.}}
\vspace*{-3mm}
\label{fig:lastcol}
\end{figure}

We caution that, among the methods mentioned above, randomization is not suitable for soft decoding,
because the magnitudes of the randomized symbols do not reflect the actual reliability level.
Therefore, in the following, we will only consider rank-1 approximation and direct method,
the BER curves of which are shown in Fig.~\ref{fig:rank1} and Fig.~\ref{fig:lastcol}, respectively.
In the performance evaluation, we consider 1) hard decision on symbols, 2) bit flipping (BF) decoding
and 3) SPA decoding. 
In some sense, hard decision shows the ``pure'' gain by incorporating code constraints. 
BF is a hard decoding algorithm that performs moderately and SPA using LLR is the best.
If we compare the SPA curves within each figure, the SNR gain is around 2 dB at BER = 1e-4.
For other curves, the gains are even larger.
On the other hand, if we compare the curves across the two figures, their performances are quite similar.
Therefore, we do not need an extra eigen-decomposition; the direct approach is just as good.

\begin{figure}[!htb]
\centering
\centerline{\includegraphics[width=8.75cm]{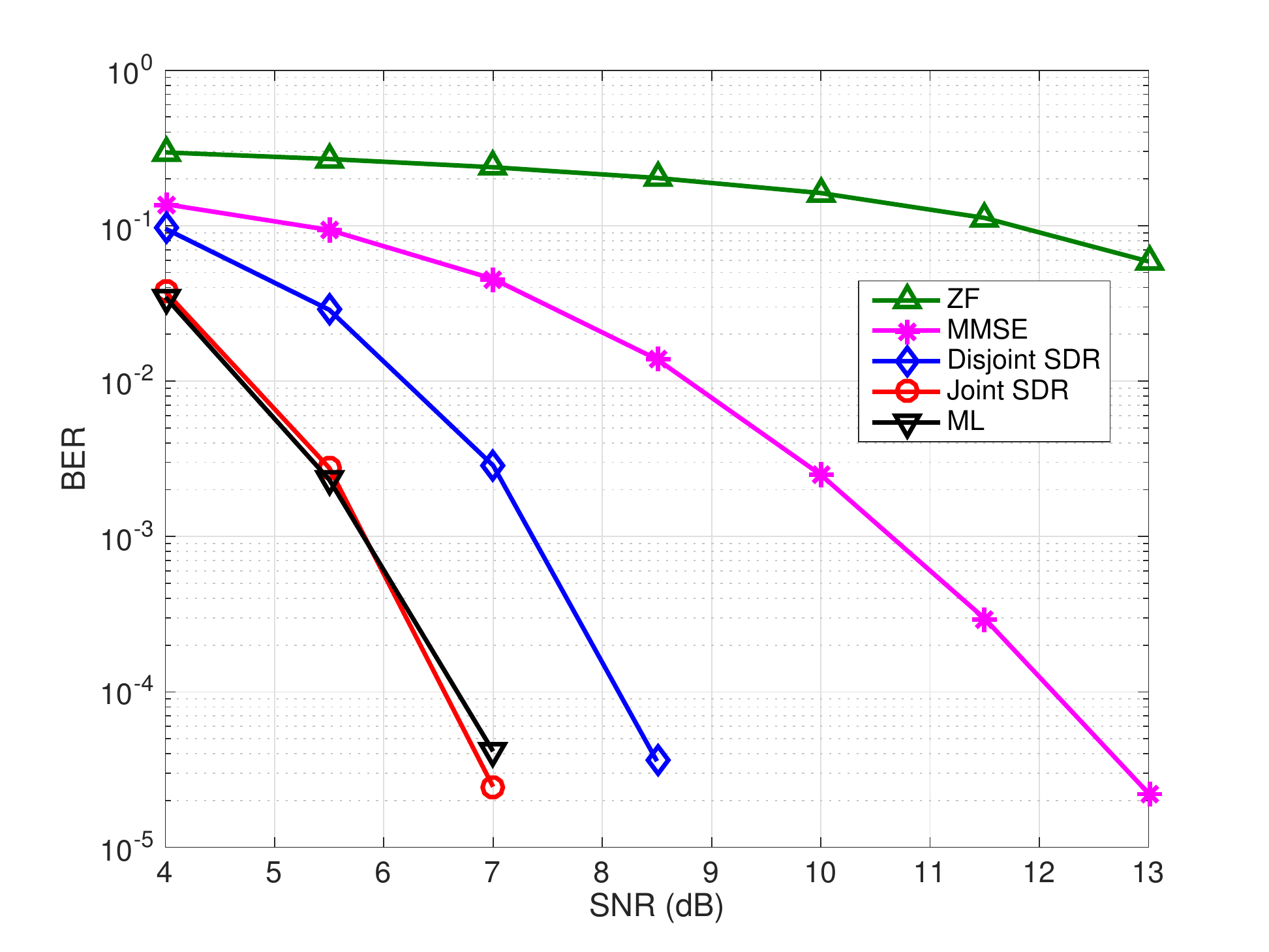}}
\caption{\small{BER comparisons after SPA: ZF, MMSE, Disjoint SDR, Joint SDR and ML.}}
\vspace*{-3mm}
\label{fig:zf_mmse}
\end{figure}

Moreover, we compare ZF and MMSE against the SDR receivers in Fig.~\ref{fig:zf_mmse}.
All BER curves are shown after SPA decoding. 
It is clear that ZF and MMSE receivers are far worse than the disjoint SDR, let alone joint SDR.
Given that ZF and MMSE are $\mathcal{O}(N_t^{3})$ complexity and SDR receiver is $\mathcal{O}(N_t^{3.5})$ complexity,
the performance gap is quite large given the relatively small difference in complexity. 
In addition, the performance of the exponential-complex ML receiver is plotted. 
Here we use a soft-output ML detector \cite{hochwald2003achieving} and then feed the LLRs to SPA decoder. 
It is seen that the BER performances of ML and joint SDR are very close, 
even though joint SDR is polynomial-complex. 

\section{Conclusion} \label{sec:con}
This work introduces joint SDR detectors integrated with code constraints for MIMO systems. 
The joint ML-SDR detector takes advantage of FEC code information in the detection procedure,
and it demonstrates significant performance gain compared to the SDR receiver without code constraints.
In current stage, this joint receiver works well with short-to-medium length FEC code. 
However, since the computation capability is ever increasing, this design should be able to accommodate longer codes.
In the meantime, we would like to conduct complexity reduction of the joint receiver in future works \cite{wang2018integrated}.
It is also interesting to investigate the robust receiver's performance against RF imperfections, 
such as I/Q imbalance and phase noise \cite{wang2017phase}.
Moreover, joint design of precoder \cite{wu2015cooperative} and receiver would be a good topic to pursue.

\bibliographystyle{IEEEtran}
\bibliography{IEEEabrv,mybibfile}
%
% <OR> manually copy in the resultant .bbl file
% set second argument of \begin to the number of references
% (used to reserve space for the reference number labels box)
%\begin{thebibliography}{1}
%
%\bibitem{IEEEhowto:kopka}
%H.~Kopka and P.~W. Daly, \emph{A Guide to \LaTeX}, 3rd~ed.\hskip 1em plus
%  0.5em minus 0.4em\relax Harlow, England: Addison-Wesley, 1999.
%
%\end{thebibliography}

% that's all folks
\end{document}